\documentstyle[12pt]{article}
\begin{document}

\title{Late-Type Dwarf 
Galaxies in the Virgo Cluster: I. The Samples}
\author{Elchanan Almoznino\thanks{email - nan@wise.tau.ac.il}
 \and Noah Brosch}
\date{{\it Wise Observatory \& School of Physics
and Astronomy,\\ Raymond and Beverly Sackler Faculty of Exact Sciences,\\
Tel-Aviv University}}
\maketitle
 
\begin{abstract}

We selected samples of late-type dwarf galaxies in the
Virgo cluster with HI information.
The galaxies were observed at the Wise-Observatory using
several broad-band and H$\alpha$ bandpasses. UV measurements were carried out
with the IUE Observatory from VILSPA, and with the FAUST shuttle-borne UV telescope. 

We describe the observations in detail, paying particular attention to the
determination of measurement errors,
and present the observational results together with published data
and far-infrared information from IRAS. The sample will be analyzed
in subsequent papers, in order to study star formation mechanisms in galaxies.

\end{abstract}


    \section{Introduction} \label{sec_int}

The process of star formation is one of the most important 
process in galactic evolution. Generally, star formation is 
initiated by the gravitational collapse of gas clouds, followed by
fragmentation into future individual
stars. When stars enter the main sequence,
 radiation pressure and stellar winds 
 push on the ambient gas and prevent it from collapsing further.

Despite extensive progress in understanding the star formation process,
there are several issues still not fully understood,
due to the complexity of this process. One can summarize the currently 
open issues,
concerning the star formation processes in galaxies, in
two major questions:
(1) What are the mechanisms that govern the star formation process,
and how do they depend on the galactic type and environment?
(2) How do the star formation rate (SFR) and the initial mass function
(IMF) depend on various galactic properties, such as
interstellar gas density,
morphology of the interstellar gas, metallicity, and
the amount of dust in the interstellar medium?

In order to find the SFR of a sample of galaxies one needs to know the IMF
characterizing the star formation in these galaxies. The IMF can be derived
by fitting a number of observed properties, such as broad-band colors, to 
a set of models of different stellar populations, with different IMFs
 (population synthesis). By comparing the various color indices to the
synthetic colors calculated from models one can, in principle, determine
the IMF of the galaxies in the sample.
This requires a database 
spanning as large a range in wavelengths as possible, to eliminate
the degeneracy of results in several conditions, being able to 
distinguish between different conditions that yield the same value for one
or more colors. It is important that one of the measured bands is in the
 UV regime, in order to trace the massive stars which contribute most of the
energy emitted in this part of the spectrum.

Using these data together with a direct massive-star tracer enables one to
draw a consistent picture of the star formation history of the sample galaxies.
One of the most common tracers of young massive stars, adopted also
 here, is based on the H$\alpha$ line intensity.
 The H$\alpha$ line has been used by many (Kennicutt 1983,
Kennicutt \& Kent 1983, Gallagher, Hunter \& Tutukov 1984, Pogge \&
Eskridge 1987, Kennicutt {\it et al.} 1994), mainly due to its high
intensity. 
The main deficiency of this method is the dust extinction.
However, the extinction is moderate in late-type irregular galaxies like
the ones used in this study, usually some 0.2--0.4$\;mag$
(van der Hulst {\it et al.} 1988).

    \section{The sample} \label{sec_SAM}

When selecting a sample of galaxies for the investigation of star formation,
one can simplify the theoretical situation 
 by selecting a sample of
galaxies of similar type, in which the star formation parameters should
basically be the same. Uncertainties due to neighborhood
influence on star formation are eliminated by selecting the
sample from a well-defined environment.

In addition, our aim is to isolate some of the star formation 
mechanisms. Small objects, such as dwarf galaxies, do not show spiral structure 
or a rotating disk which may complicate the star formation triggering parameters. 
Therefore they make good candidates for our research. 
 
We selected a sample 
of late-type dwarf galaxies, all in the Virgo cluster area. 
This is a nearby region, thus
the galaxies appear reasonably bright, and are at a high
Galactic latitude ($63^\circ$ \raisebox{-.6ex}{$\stackrel{<}{\sim}$} $b$
 \raisebox{-0.6ex}{$\stackrel{<}{\sim}$} $80^\circ$)
where the extinction is 
very small. In addition, as the galaxies are members of a rich cluster, 
the sample enables testing for the effect of cluster environment on the star 
formation properties.

          Binggeli, Sandage \& Tammann (1985, BST) catalogued more than 2000
          dwarf galaxies in the Virgo cluster. Most objects are dwarf
          ellipticals, but a few hundred are late-type dwarf galaxies.
          Hoffman {\it et al.} (1987) and Hoffman {\it et al.}
          (1989) studied all late-type dwarfs in the BST sample with the
    Arecibo radio telescope and produced single-beam HI flux integrals and
          shapes of the 21 cm line. Our sample of late-type dwarfs was
          extracted from their list of objects with non-zero HI measurements, using the 
following selection criteria:

Two subsamples of dwarf galaxies were selected,
dichotomized by their total HI content (high or low). Then, in order to test the
dependence of star formation properties on the galactic type, we
separated between high surface brightness (BCD and ImIII/BCD)
and low surface brightness galaxies (ImIII--ImV). 
 We, therefore, constructed four subsamples of late-type dwarfs
from the BST catalog (hereafter Virgo Cluster Catalog - VCC).
 
 The selections criteria for the four subsamples were
chosen in order to have similar numbers of objects in each group,
and are characterized as follows:\\
\\
\begin{tabular}[b]{|c|c|c|c|}
\hline
Name & Type & $F_I$ (mJy km/s) & No. of \\
 & & & objects \\
\hline
 High/high (H/H) & BCD, ImIII/BCD & $F_I > 1500$ & 9  \\
 High/low (H/L) & BCD, ImIII/BCD &  $0 < F_I < 500 $ & 7 \\
 Low/high (L/H) & ImIV, ImIV/ImV, ImV & $F_I > 1000 $ & 4 \\
 Low/low (L/L) & ImIII/ImIV, ImIV, ImIV/ImV, ImV & $0 < F_I < 600 $ & 5 \\
\hline
\end{tabular}

\begin{figure}[htbp]
\vspace{8.6cm}
\includegraphics{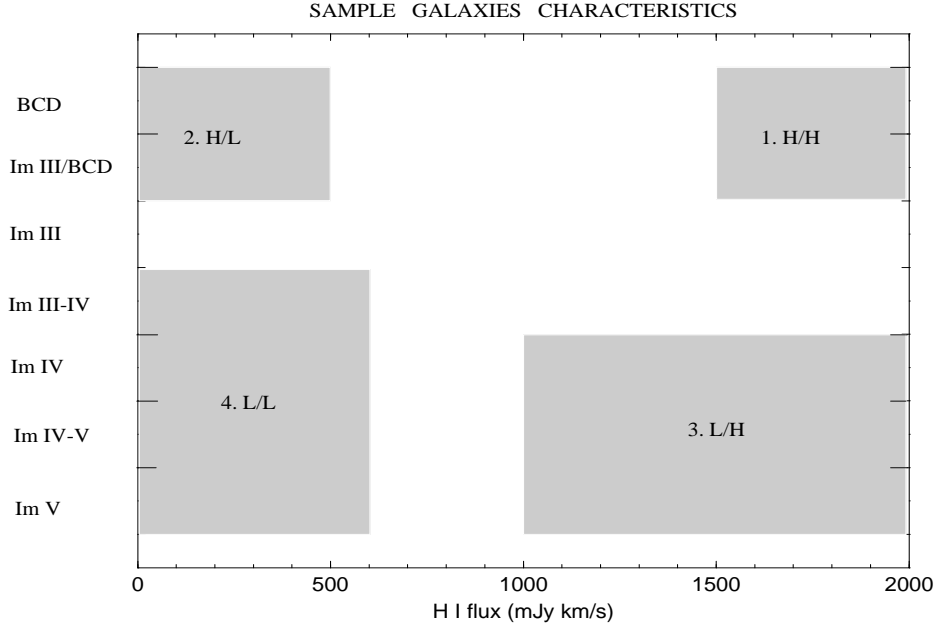}
\caption{\protect \footnotesize{Distribution of the four subsamples
of dwarf galaxies in the morphological type - 
total HI content plane. The designations H/H, H/L, L/H and L/L are explained
in the text. Some "high HI" members may have
 $F_I > 2000 \;mJy km/s$.}}
\label{fig_SAM}
\end{figure}

%
\begin{table}[tp]
\vspace{12.1cm}
\includegraphics{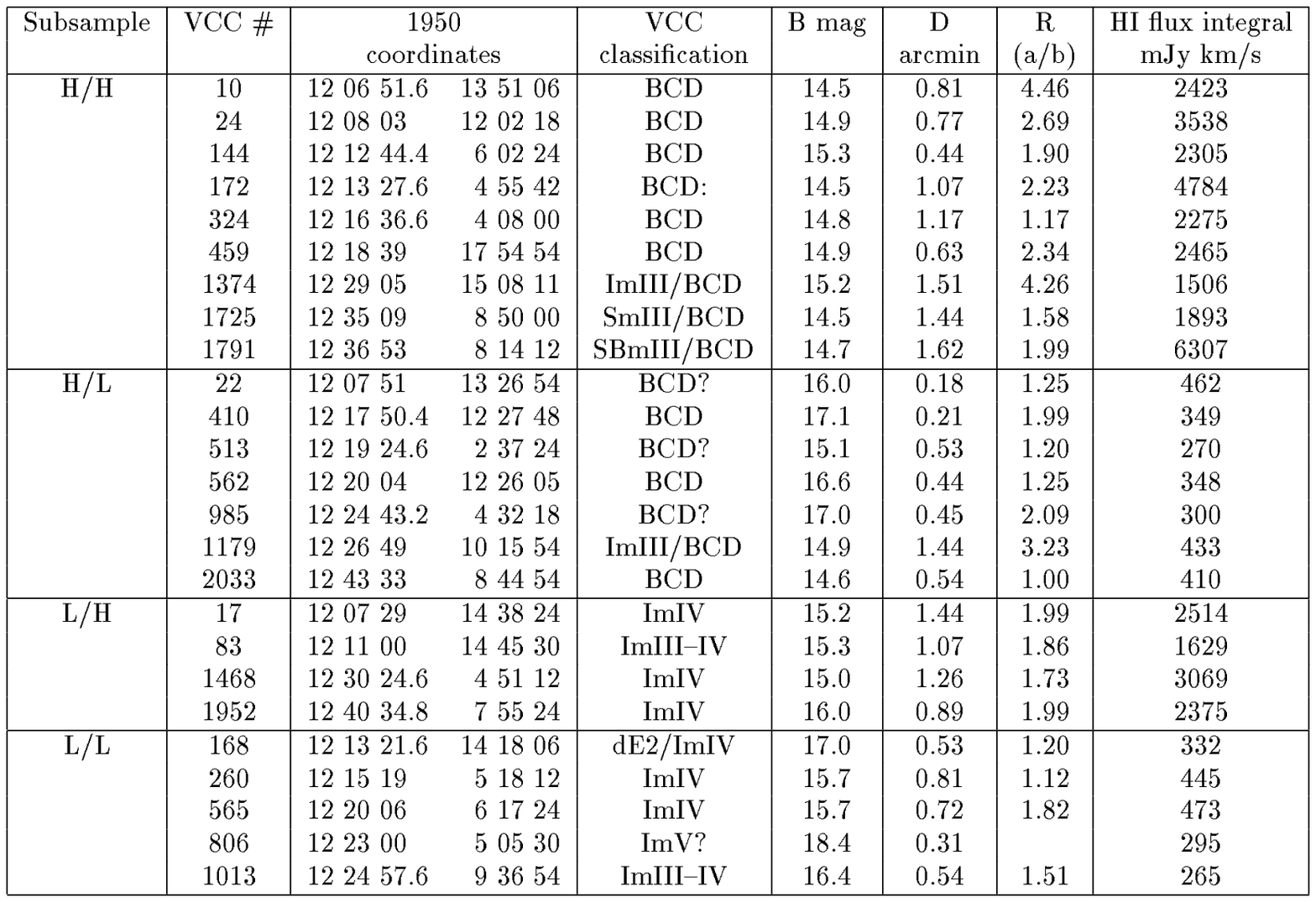}
\caption{\protect \footnotesize{The sample of dwarf galaxies, from the
VCC, studied here. The major axis D is in arcminutes, and the ratio 
between the major and minor axes is denoted by R. The HI flux integrals 
are from
Hoffman {\it et al.} (1987) and Hoffman {\it et al.} (1989).}} \label{tab_sam}

\end{table}


The selection criteria are schematically displayed in Fig.~\ref{fig_SAM}, and
the sample galaxies, extracted from the VCC using these criteria,
 are presented in Table~\ref{tab_sam}. The mass of neutral hydrogen in
each galaxy is directly related to its HI flux integral, and will be presented
in~\ref{sec_Addata}.

\begin{figure}[htbp]
\vspace{12.4cm}
\includegraphics{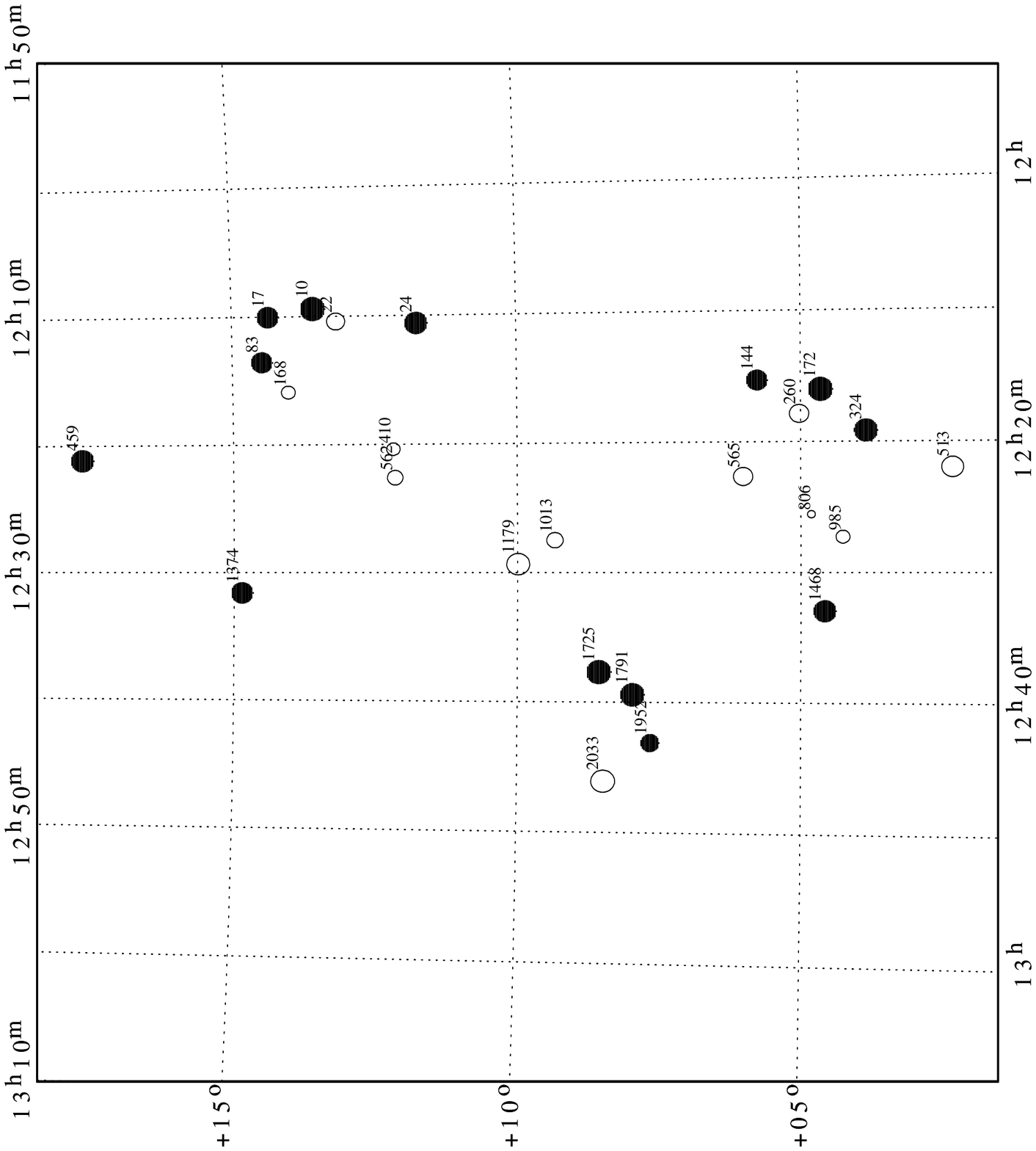}
\caption{\protect \footnotesize{The sample galaxies projected on the
 sky. Each galaxy is labeled 
by its VCC
number and the size of the circles indicates its B magnitude from the
VCC. High HI
galaxies are symbolized by filled circles and low HI galaxies by open 
circles.}}
\label{fig_Vsam}
\end{figure}

The intention in the dichotomy in both galactic type and HI content is
to attempt to discover even weak dependence of star formation on these
parameters.
We expect to find a difference in the observed properties which measure
recent and present star formation, such as
broad-band colors and H$\alpha$ line strength, between the low and high
surface brightness subsamples. We are aware of the small number of objects in 
each subsample, a result of observational necessity.

The spatial distribution of the sample galaxies is displayed in 
Fig.~\ref{fig_Vsam}. Low HI galaxies are represented by open 
circles, high HI galaxies by filled circles. High
HI galaxies are missing from the central region of the cluster, while
low HI galaxies appear to be evenly distributed within
 the cluster. Although our sample
is small, it supports the hypothesis of gas stripping off near the cluster
core.

   \section{Data collection} \label{sec_Opt}

Our data collection consists mainly of optical observations.
In addition, UV observations and published information are used and
 presented here.
 
    All optical observations were carried out at the Wise Observatory
    (WiseObs) in Mizpe-Ramon, Israel.
    A few of the sample galaxies were observed by NB in spring 1988.
    During this run the galaxies were imaged using the observatory's RCA-CCD
    camera. Each galaxy was measured for 10 min. through a standard V
    filter.

    The other observations were carried out from 1991
    to 1993. 
These observations
     included three types of measurements: galaxy imaging in broad bands,
photometry of reference stars, and galaxy imaging in H$\alpha$.

    \subsection{Broad-band imaging} \label{sub_Bband}

    Each galaxy was imaged through B, V, R and I filters with
    exposure times which varied between 5 and 30 min. The RCA CCD was used
    in most cases, except for some runs 
in which the FOSC was used in imaging mode.
For each object a
    reference star close enough to appear in the galaxy's CCD
    frame was selected and was imaged together with the galaxy. The
RCA CCD is a 512x320 pixel array, thinned and back-illuminated, with $30 \mu m
= 0".9$ pixels. The FOSC is a focal reducer device with a TI $1024^2$ pixels,
where each projected pixel is 0".7.

    \subsection{Photometry of reference stars} \label{sub_Refstar}

    In order to put the flux on an absolute scale, the reference stars 
   imaged with the galaxies were measured with the Two-Channel
Photometer of WiseObs. We used only channel 1,
 with a Hamamatsu RCA C31034A
    photomultiplier type with a GaAs cathode in a 
thermoelectrically cooled enclosure.

    The reference stars were chosen so that their magnitudes were in the 
proper range for the photometer: 10--16$\;mag$.
 Each of the reference stars was measured through 
    B, V, R and I filters together with photometric standard stars from the
    Landolt equatorial sequences (Landolt 1973, 1992).

    The photometry of the reference stars requires optimal atmospheric
    conditions, as only then can the atmospheric extinction be
    accurately determined. For this reason, these measurements were repeated
    during several nights in order to obtain more reliable results
    and test their consistency.

    \subsection{H$\alpha$ measurements} \label{sub_Ha}

    Most of the sample galaxies were imaged through narrow-band filters to
    derive their H$\alpha$ emission. For each galaxy two filters
    were used: one which contains the H$\alpha$ line ("H$\alpha$ filter"),
    and the other which samples the continuum radiation adjacent to it towards
    longer wavelengths ("off-H$\alpha$").

    One of the three H$\alpha$ filters listed in Table~\ref{tab_fil}, 
    which differ by their central
    wavelength, was used to image each galaxy according to its redshift, while
    the same off-H$\alpha$ filter was used for all galaxies. The central
    wavelength and width of these filters were chosen with the following
    requirements:

    \begin{enumerate}

    \item{There is no overlap between the off-H$\alpha$ and any of the
    H$\alpha$ filters, while this bandpass is still close enough to the 
line for sampling directly the continuum radiation,}

    \item{There are no significant spectral lines in the bandpass of
    the off-H$\alpha$ filter,}

    \item{The H$\alpha$ filters cover the range of velocities of the Virgo
    cluster ($\sim$ 0 -- 2500 km/s), while any H$\alpha$ line emitted
    within this velocity range
    falls in one of the filters' bandpass within at least 90\% of its peak
    transmission.}

    \end{enumerate}



\begin{table}[h]
    \begin{tabular}[b]{|c|c|c|c|}
\hline
    Filter ID. & $\lambda_0$ & $\Delta \lambda$ (FWHM) & Peak
transmission\\
\hline
    H$\alpha-2$ & 6562\AA & 50\AA & 67.5\% \\
    H$\alpha-3$ & 6586\AA & 48\AA & 66.0\% \\
    H$\alpha-4$ & 6610\AA & 55\AA & 69.7\% \\
    off-H$\alpha$ & 6700\AA & 53\AA & 70.0\% \\
\hline
    \end{tabular}
\caption{\protect \footnotesize{The main characteristics of the
H$\alpha$ filters used at WiseObs.}}
\label{tab_fil}

\end{table}

One should note the [SII] lines (6716\AA\ and 6731\AA) which may
be sampled by the continuum filter. The contribution of these lines is
expected to be small in these galaxies, at the most $\sim$15\% of the H$\alpha$ radiation 
(e.g. Gallego {\it et al.} 1996). The shorter of these lines is reduced by the filter's 
response by $\sim$10\% while the other is reduced by $\sim$50\% for zero redshift systems.
For higher redshift galaxies these lines are extinguished further, and therefore
we can ignore their contribution in our sample of galaxies.
The filter characteristics described above are for light falling normal to 
the plane of the filters, and
they vary, in principle, with the incidence angle.
Since the observations were carried out
with the CCD placed in the focal plane of the f/7 converging
beam of the telescope, the incidence angle on the filters
is between $1\!^\circ\!.8$ -- $4\!^\circ$. 
A measurement at the Tel-Aviv University Applied Physics
Department showed that the central wavelength of
the three H$\alpha$ filters shifts by 1.5\AA\ to the blue relative
to the nominal wavelength due to this effect, while the
off-H$\alpha$ shifts by 3.8\AA\ blueward. We conclude that
these changes are not significant for our measurements.

    For each of the galaxies, at least two 20 min exposures were obtained
    through each of the two filters.
In addition, a spectrophotometric standard star
    (HZ44) was measured several times each night through different air
    masses with the same filters, in order to derive the
 atmospheric extinction
    and absolute photometric calibration for the H$\alpha$ images.

\subsection{UV observations and data} \label{sec_UVobs}

Seven of the sample galaxies were observed with the IUE, two of them in 1986
within the US program EGITT. 
The other five were observed by EA
in 1991 and 1993. The observations took place at the VILSPA tracking station,
Spain, and each galaxy was observed for a full VILSPA observing shift,
 $\sim$6 hours of integration. In all cases the spectrum was
 obtained through the IUE large aperture,
which is a $\sim10"\times20"$ oval, using the short wavelength camera (SWP) 
- low dispersion setup, which covers the spectral range of 1200--1950\AA.

In late March 1992 the ATLAS-I space shuttle mission took place,
during which UV observations were carried out with the shuttle-borne
telescope FAUST ({\it Fus\'{e}e Astronomique pour l'Ultraviolet
STellaire} or {\it FAr Ultraviolet Space Telescope}). 
This UV imaging
instrument was developed jointly by the Laboratoire 
d'Astrophysique Spatiale of the CNES, and the Space Astrophysics Group at Berkeley.
The FAUST UV bandpass has $\bar{\lambda}\simeq 1650$\AA\  with $\Delta
\lambda \simeq 400$ \AA. 
During the ATLAS-I mission it imaged a number
of $\sim8^{\circ}$ wide sky areas, among which are three overlapping frames of the 
Virgo cluster (Brosch {\it et al.} 1997a).

These FAUST frames were used to obtain UV
data for the dwarf galaxies in our sample.

%
%

\subsection{Additional data} \label{sec_Addata}

A few of the sample galaxies were detected by IRAS and appear in the
IRAS point source catalog.  
In addition, co-addition of the individual IRAS scans for a deeper detection
or an upper flux limit
was done by NB for the objects which do not appear in the IRAS catalog.

For every galaxy of the sample, except for VCC806 and VCC1013 where 
observations were not possible due to software problems, 
we have, thus, the IRAS data, either as
absolute fluxes or as upper limits of flux. In addition, some objects
are FIR sources listed in the IRAS Faint Source Catalog. Their flux densities
together with the other galaxies data are listed in Table~\ref{tab_misc2},
but without errors. The typical error of the FSC, for faint IRAS sources,
is $\sim$15\% of the listed flux density.

Our sample uses results from Hoffman
{\it et al.}
(1987) and Hoffman {\it et al.} (1989) obtained with the Arecibo radio telescope.
From these, we calculated the HI mass following the 
prescription in the Third Reference Catalog of Bright Galaxies
(de Vaucouleurs {\it et al.} 1991), as: 
\begin{equation} \label{e_MHI}
M_{HI} = 235.5 \;S_{HI}\; D^2
\end{equation}
where $M_{HI}$ is in M$_\odot$, $S_{HI}$ is the HI line flux integral in
mJy km/s  and D is the distance to the object in Mpc. 

 In order to derive the HI mass of these galaxies, a common 
distance is adopted for all objects.
A value of 18 Mpc may well represent
the average distance of galaxies in the Virgo cluster (e.g., Fouqu\'{e} {\it 
et al.} 1990), and is the value adopted here. No
more accurate information is available on the distance to the type of
galaxies discussed here
(such as a Tully-Fisher relation), and in any case, such a relation
is usually very dispersed for irregular galaxies and the uncertainties are
large. 
We adopt, therefore, a distance of 18 Mpc and bear in mind that
the uncertainty in this distance can be up to $\sim$15\%.
These values, together with the IRAS
data, are given in Table~\ref{tab_misc2}.

%
\begin{table}[htbp]
\vspace{13.5cm}
\includegraphics{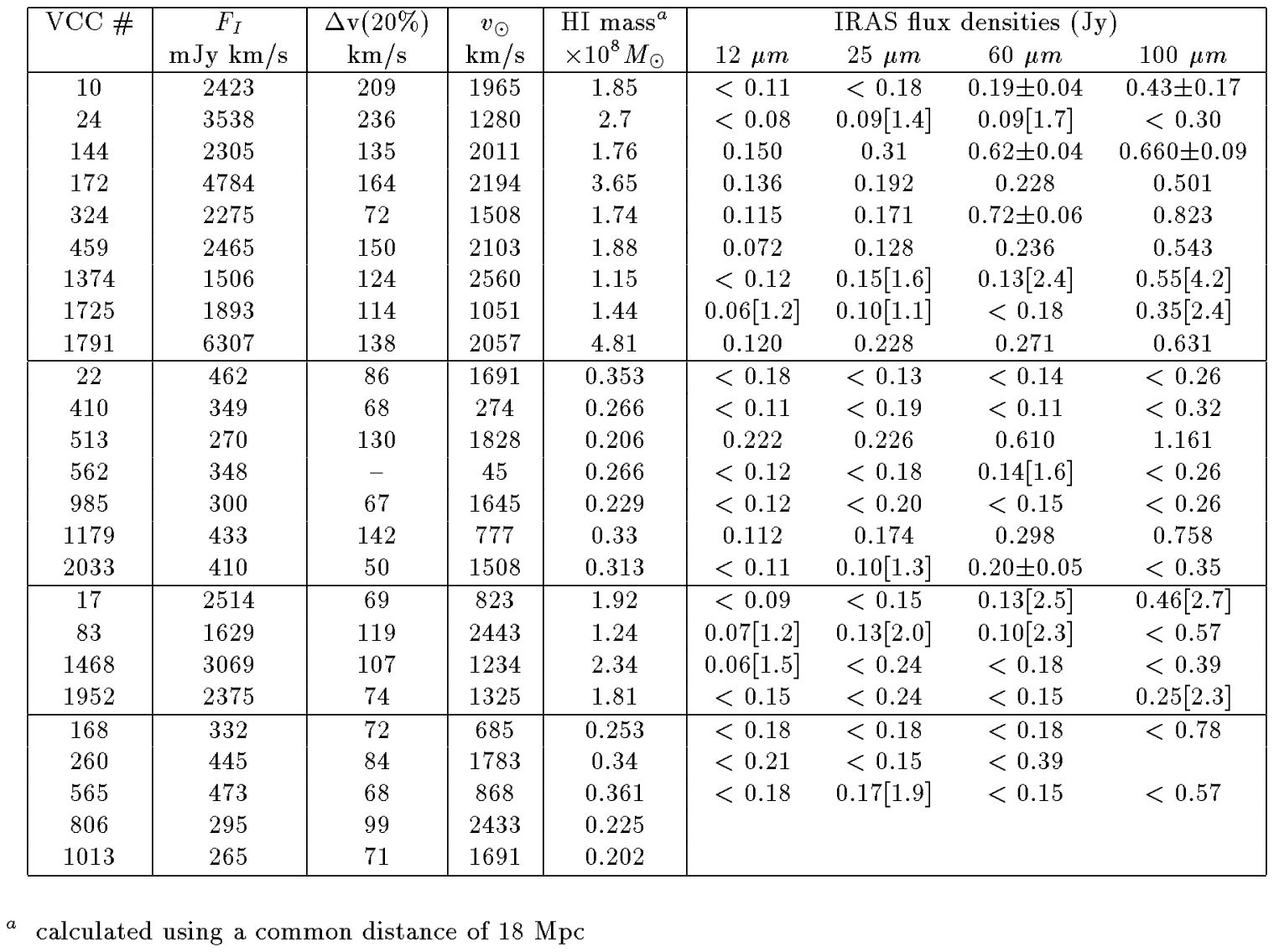}
\caption{\protect \footnotesize{IRAS and radio data of the VCC sample 
galaxies}} \label{tab_misc2}

\end{table}


\section{Data reduction} \label{sec_reduc}

  All data reduction was performed at WiseObs headquarters,
Tel-Aviv University. The CCD images were reduced
 with the VISTA image-processing software package,
 (Pogge {\it et al.} 1988), using normal procedures. 
Photometric data was reduced with the observatory's photometry package.

\subsection{Broad-band colors} \label{sub_BBreduc}


    The photometer data were reduced with a composite extinction-transformation
    program. This program computes instrumental magnitudes and colors and
    fits them by least-squares to the catalog values of the standard stars.
    The fit is linear in air mass and second-order in colors, and it takes care
of effects like the variation of extinction across the bandpass, as well as 
differences between the filters used and the Johnson-Kron-Cousins standard 
bandpasses.

    As mentioned before, the photometric observations were repeated during
    several nights for most of the reference stars. Each night was ranked
    according to the goodness of fit of the standard stars parameters.
    The results of the different nights were consistent with each other, 
 and for each reference star the result from the
    'best' night was adopted.

   The broad-band images of every sample object were
merged to produce a final frame with the best S/N
    available in each band. 
The object's total intensity was measured on each final frame, together with
    the total intensity of the corresponding reference star.
Since the objects are irregularly shaped, no model fitting to their
shape could usually
be done. Therefore, an empirical way was adopted, as follows:

    A polygonal area, which contains the entire object as determined by
visually inspecting the image, is constructed
    manually upon the displayed image, using the deepest of the four
    available images of that object. The guideline for setting the polygon
border is that the signal-to-noise ratio for each pixel of the object's
image falls well below unity,
while trying not to include too much of the sky noise. This 
corresponds to a typical surface brightness of $\sim 27\;mag/\Box"$ in V,
depending on the total exposure time and sky background level.

\begin{figure}[htbp]
\vspace{9.1cm}
\includegraphics{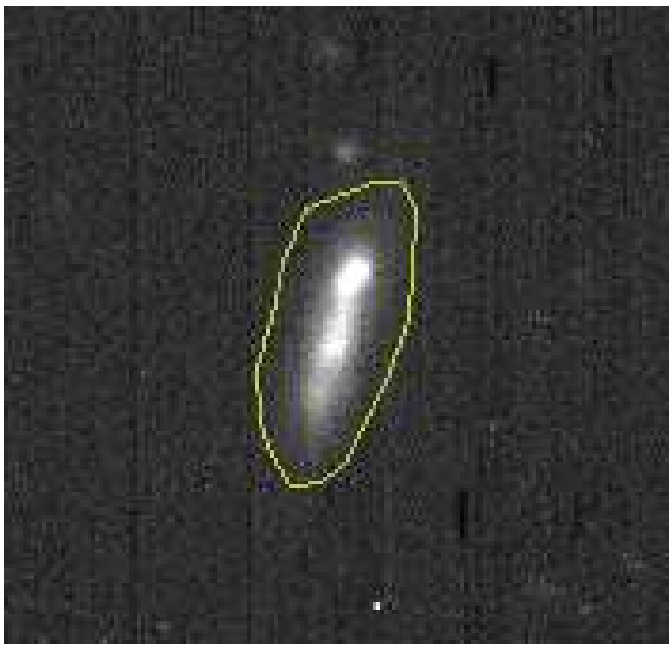}
\caption{\protect \footnotesize{The image of VCC1374, with the surrounding 
polygon used for measurement.}}
\label{fig_pol}
\end{figure}

The total counts within this region are
    summed-up and the appropriate sky background 
is subtracted. The sky value and $\sigma_{sky}$ (standard deviation of a 
pixel's value from the sky level) were calculated using the SKYALL 
procedure (Almoznino {\it et al.} 1993), which
adopts the most probable intensity value as the sky level.
The error of the object intensity $I$ is given by:
\begin{equation} \label{e_err1}
\Delta I = \sqrt{\frac{I}{G} + 2P\sigma_{sky}^2}
\end{equation}
where G is the gain of the instrument, and $P$ is the number of pixels within the 
polygon.
An example of a galaxy image with the polygon used for its measurement
is shown in Fig.~\ref{fig_pol}.

The reference star was measured in the same way as the object, only within a
$D=\sim 20$ arcsec
circular 'aperture' centered on the star, instead of a polygon. 

With the photometrically measured magnitudes of the reference star in each band,
 the corresponding magnitudes of the object are obtained directly
from the CCD frames from the ratio of intensities. Actually, the
derived quantities were the V magnitude and the color indices
    B--V, V--R and
R--I. The advantage of this method is that observations of the sample galaxies,
which consume large amounts of telescope time, do not require
any special observing conditions, since both the object 
and the reference star appear
on the same CCD frame, and the measurement is relative. 
An effect that may be influencing
the results, is the possible difference between the effective response of the
 CCD+filters observing system and the standard Johnson-Kron-Cousins system.
 The implication of this effect on the accuracy of the rersults
 is discussed in~\ref{sec_res}.

The object V magnitude is given by:
\begin{equation} \label{e_magv}
m_v = -2.5 \; \log \left ( \frac{I_o}{I_s}\right ) + V_{phtm}
\end{equation}
where $I_o$ is the object CCD intensity, $I_s$ is the reference star CCD
intensity, and $V_{phtm}$ is the reference star V$\;mag$  from the 
photometric measurement.
The error of the derived magnitude is:
\begin{equation} \label{e_errmv}
\Delta m_v = \sqrt{1.0857^2(\delta_o^2 + \delta_s^2) + \Delta_{phtm}^2}
\end{equation}
where $1.0857=2.5\;log_{10}e$, $\delta_o=\frac{\Delta I_o}{I_o}$ and $\delta_s
=\frac{\Delta I_s}{I_s}$ are the relative
errors of the object and star CCD intensity, and $\Delta_{phtm}$ is the error
of the photometric result.

The object color indices and their errors are derived in the same way as the V
magnitude, except that intensity ratios between two CCD frames are used,
instead of intensities. 
In a single case (VCC 1468) the I band image had a very intense and 
nonuniform sky background relative to the low object signal, causing an
 unreliable R--I result of \mbox{--0.25.} This
result is discarded here, while other colors of the same galaxy are used.

    In addition to broad-band magnitudes and colors, approximate
    monochromatic magnitudes at $0.44\ \mu m$, $0.55\ \mu m$, $0.64\ \mu m$ and
    $0.79\ \mu m$ were obtained from the broad-band data. This was done using
    the corresponding data for flux densities at the central wavelength of B,
 V, R and I broad-bands for a normal zero-magnitude 
star from Bessel (1979). The flux density for a galaxy at a wavelength that 
corresponds to a certain band $i$ is given by 
$F_{g,i}=F_{s,i}\times 10^{-0.4\ m_{g,i}}$, where $F_{s,i}$ is the flux
density of the zero-magnitude star and $m_{g,i}$ is the broad-band
magnitude of the galaxy at band $i$.
The flux densities are converted
to monochromatic magnitudes by: \mbox{$m_\lambda = -2.5 \log
    [F_\lambda $(erg/s/cm$^2$/\AA)$] - 21.175.$} The errors are the same as
the corresponding broad-band errors. 

    \subsection{H$\alpha$ flux measurements} \label{sub_Hareduc}

The purpose of observing the galaxies in H$\alpha$ is to measure 
the line emission from each galaxy. For this, one should subtract the
continuum radiation from the flux measured through the H$\alpha$ filter. 
This is done by observing each galaxy through two filters: one
centered on the line (H$\alpha$ filter) and another rejecting 
emission lines and (presumably) sampling the continuum (off-H$\alpha$).
The procedure entails the determination of the
flux {\it density} of the continuum (sampled by the off-H$\alpha$ filter),
scaled to the transmission profile of the H$\alpha$ filter, and its
subtraction from the flux measured by the latter filter.

In order to derive the absolute fluxes of the objects, the atmospheric
extinction should be compensated. This is done
using the results for the spectrophotometric
standard star, which was measured through different air masses during
the same night of observation. Naturally, this type of reduction uses 
magnitudes rather than fluxes. Only after the atmospheric extinction
is taken care of, can fluxes be derived.

    The  H$\alpha$ data consist of at least two frames in the H$\alpha$
filter and two in the off-H$\alpha$ filter, for each of the sample galaxies.
In contrast to the broad-band data where frames were combined,
here each frame was reduced to
yield a separate monochromatic magnitude for the object.

Each of the galaxies was measured inside the same polygonal area used for
its broad-band image. The method of measurement was basically the same as
for the broad-band frames,
with the exception  that the sky background level was computed several 
times using different sky
'boxes' around the objects. This was because the signal from the galaxies was
    very weak in a 50\AA\ narrow band, and slightly different sky values
 make a difference in the results. The mean of these results
    was adopted as the object counts, while the error was usually taken as
    the largest of the errors of each measurement. The standard deviation of
    the results was normally smaller than this, but whenever it was larger -
    it was adopted as the error. In some cases the result was consistent
    with zero, mostly in the off-H$\alpha$ band. The error of the galactic
counts was calculated by equation~\ref{e_err1}.

The photometric standard HZ44 was measured through a 'circular
    aperture' on each of its frames.
We converted the results for atmospheric extinction and put them on the monochromatic
magnitude scale of HZ44  (Massey {\it et al.} 1988), where
the stellar spectrum is convolved with 50\AA\ wide bands, separated by
50\AA. 
The following results were used for HZ44: \ 
$m_{6562}=12.20,\ m_{6586}=12.16,\ m_{6610}=12.12,
\ m_{6700}=12.16. $

A mean flux result was calculated for each object in each of the two bands,
using the available monochromatic magnitude results for this band. 
The total flux obtained through any of the two filters is the 
average flux
density in the appropriate wavelength times the effective filter width. 
The shape of the transmission curves of our narrow-band filters indicates that
their FWHM represents well their effective width, thus,
 the FWHM of the filters listed in~\ref{sub_Ha} was adopted
as their effective width.
 
To obtain the net H$\alpha$ line flux, the continuum contribution
is subtracted from the total H$\alpha$ band flux. 
However, the result of the off-H$\alpha$ filter may not represent
the exact amount of continuum radiation at the H$\alpha$ band, as the
continuum flux itself gradually changes with wavelength.  This color term is
small, but may be significant in the case of H$\alpha$ faint galaxies.
To compensate for this, the rough SED from the broad-band data was
considered (its derivation is described in~\ref{sub_com}). 
The monochromatic magnitudes at $0.55\;\mu m$ and
    $0.79\;\mu m$, derived from the V and I magnitudes, were interpolated 
linearly to yield a rough, gradual change across the spectral region
between them. This information was used to obtain the expected
change in continuum
radiation between the two H$\alpha$ narrow bands. The R band magnitude was
not used since it contains the H$\alpha$ line and may be misleading.

The SEDs of the galaxies are displayed in 
Figs.~\ref{fig_sed1} and~\ref{fig_sed2}.
For all the sample galaxies the SED decreases with increasing wavelength
and the relative change of the continuum radiation between the off-H$\alpha$
and H$\alpha$ bands is
from 0 to 2.5\%. For the H$\alpha$-bright galaxies the H$\alpha$ 
flux may be as high as twice that of the off-H$\alpha$ band and clearly
the effect of continuum sloping is insignificant,
as the accuracy of our measurements is of order of 5\%. For
H$\alpha$-faint galaxies, however, taking this effect into account can
change our interpretation for a certain object,
from a significant H$\alpha$ flux to $\sim$zero H$\alpha$ flux.

If we denote the expected ratio between the H$\alpha$ band {\it continuum}
flux and the off-H$\alpha$ flux, as derived by the above procedure, by $R$, 
the net H$\alpha$ line flux is given by:

    \begin{equation} \label{e_Haflux}
    F(H\alpha) = (F_{line} - R\times F_{off}) W_{line}
    \end{equation}
    where $ F_{line} $ and $ F_{off} $ are the average flux
    densities obtained through the two bandpasses, and
    $W_{line}$ is the FWHM of the H$\alpha$ filter. 
The H$\alpha$ equivalent width is given by:

    \begin{equation} \label{e_Haeqw}
 EW[H\alpha] = \frac{F(H\alpha)}{R\times F_{off}} W_{line}
    \end{equation}

The errors were calculated by standard procedures from the errors of
 $ F_{line} $ and $ F_{off} $ while the parameters 
$R$ and $W_{line}$ are taken without error. This yields a relative error
of the H$\alpha$ flux from $\sim$5\% in the H$\alpha$ bright galaxies
to more than unity in the galaxies with non detected line radiation.
Note that all the H$\alpha$ fluxes and equivalent widths
derived here are actually H$\alpha$ + [N II] fluxes and equivalent widths, 
as the filters used include also the accompanying
nitrogen $\lambda \lambda$ 6548, 6583\AA\ lines. This will be discussed later 
in more detail.

Using the total flux, the H$\alpha$ flux per unit solid angle was derived for 
each galaxy. The angular size of the galaxies was determined by constructing
a confining polygon around the galaxy, with its border 
set to a S/N $\sim1$ (smaller than the one used for the measurement of total
fluxes). The border was set on 
the deepest image of each
object and it corresponds to $\sim25\;mag/\Box"$ in B.

    \subsection{UV data reduction and combination with optical data}
    \label{sub_UVred}

As mentioned in section~\ref{sec_UVobs}, two types of UV data were 
collected for the sample
galaxies: IUE spectral data and FAUST imaging data. Our aim is to construct
a SED for as many galaxies as possible from the sample. For this, a
derivation of UV monochromatic magnitudes was performed at WiseObs.

For the five galaxies observed during 1991 and 1993 the spectra were
re-derived from the photometrically corrected IUE images. 
Reseau marks and 'hot pixels' were interpolated over, to obtain more 
reliable spectra of the objects.
    The spectra of all the galaxies were faint and noisy, and no spectral
    lines were clearly visible. To obtain the continuum flux, the spectra
were convolved with a 30\AA \ wide Gaussian kernel, and only the smoothed spectra
    were considered. This was performed also for the two sample
    galaxies observed in 1986, where the Uniform Low-Dispersion 
archive spectra were used.

    The smoothed spectra were measured in two 100\AA\ wide
    'bandpasses', centered at 1350\AA\ and 1850\AA. The error of these
    measurements was estimated to be between 15\% and 20\% according to
the noise at each band. For two objects (VCC562 and VCC2033) this
procedure yields $\sim$zero flux, and their spectra were not reduced
further. A conversion was
    made from $F_\lambda$ to monochromatic magnitudes with 
$m_\lambda = -2.5\; log [F_\lambda (erg/s/cm^2/$\AA$)] - 21.175$.  
The error was calculated as
    in~\ref{sub_Hareduc}.

    The IUE large aperture is $\sim\!10"\times20"$ in size. Therefore, the
    spectrum we obtained does not represent the radiation emitted by the
    entire galaxies, but rather the energy from their bright regions. However,
    because the galaxies in our sample are small in size, most or
 all of the radiation comes from a region of that size.

    In order to ensure that no bias is present in the results, the optical
    broad-band images of the galaxies with IUE data were
    sampled through 'ovals' as similar as possible to the IUE aperture, in
    size, shape and orientation during the IUE exposure. 
The reduction was
    the same as for the broad-band photometry from CCD images, with the exception 
that the area of sampling was
    different. Monochromatic magnitudes in the optical were derived for the
    central regions of these galaxies to produce a meaningful combination
    with the UV data and a reliable SED of the galaxies.

    The above treatment was not performed 
 for VCC324=MRK49, which is very large for IUE ($\sim40"$). Because of its
    size, a small change in the location, orientation or size of the
    artificial oval used for sampling, may lead to great changes of the final
result, which, in turn, may lead to an error in the derivation of the SED.
Therefore, the UV data
for this galaxy was discarded.

    As mentioned above, three fields in the Virgo cluster were imaged by 
the FAUST instrument.
    Each field is about $8^{\circ} $ in diameter and together they cover
most of the cluster. 
A merged picture of the three images
is shown in 
Brosch {\it et al.} (1997a), along with a full analysis of the FAUST 
observations in the Virgo region. 
Note also that
a general catalog of all FAUST sources was produced by automatic correlations
against existing catalogs (Bowyer {\it et al.} 1995).

The FAUST images underwent flat-fielding corrections at
Berkeley, to obtain frames in calibrated units
of FAUST-counts per second. Each frame was compared to a predicted UV
map of the corresponding area of the sky, derived from the SAO catalog
(Shemi {\it et al.} 1993), in order to identify the SAO stars in it.
These were used to derive the transformation from the
image pixel coordinates to celestial coordinates. About 15 stars were 
used in each
frame, and the standard deviation of the calculated coordinates from the
real ones was $\simeq 0'.5$ (0.45 pixel).

The expected location of every sample object was found on the frames
using these transformations.
At each location
'aperture photometry' was performed in much the same
method as for the reference star measurement in~\ref{sub_BBreduc},
with the exception that the error calculation did not take the 
photon-statistics
error into account. This is because in this type of frames the sky noise
dominates the error of the measurement in comparison with other effects.

Two of the sample galaxies were reliably detected on the FAUST
frames: VCC1725 and VCC1791. For the other objects, an upper
limit for the UV radiation was derived as follows: whenever an object 
was expected
to appear in more than one FAUST frame, its location was measured 
separately on
each frame. Results for objects close to the image edges were ignored,
due to very low S/N, and a $1\sigma$ upper limit was derived whenever
the accuracy was worse than 30\%. For objects which had more than one
acceptable result, the `best' result was adopted,  i.e., the lowest upper
limit value.
The results were then transformed to approximate monochromatic magnitudes
at 1650\AA\, using the calibration from Brosch {\it et al.} (1995):

\begin{equation} \label{e_FA}
 m_{_{FAUST}} = m_{inst} + 11.62
\end{equation}
where $m_{inst}$ and its error are the same as in~\ref{sub_Hareduc}.

Unfortunately, the two objects which were detected appeared
only on one frame, so no cross check for galaxy fluxes was possible. 
However, cross checks
between the three frames were carried
out using other objects which appear in two, or three
images, yielding good agreement between their different count results:
disregarding very bright objects, in which case distortions arise, 90\%
of the objects have values that differ by less than 1.5$\sigma$ 
from each other.
It should be noted that only two galaxies were both measured by IUE 
{\it and} are within the field of view of FAUST. These are VCC562 and VCC2033,
which have FAUST UV upper limits consistent with the \mbox{S/N $\ll 1$}

    \section{Results} \label{sec_res}

    \subsection{Broad-band magnitudes and colors} \label{sub_br}

    In general, the V magnitude of the galaxies lies within a range of
   $\sim0.6\;mag$
    of their listed blue magnitude from the VCC. Caution should
be exercised when comparing our results with other published data, since most
of the photometric observations of these galaxies were carried out through 
apertures with various diameters, between 19" and 120" (Hunter \& Gallagher
1985, Gallagher \& Hunter 1986, Gallagher \& Hunter 1989, Drinkwater
\& Hardy 1991). We compared our broad-band results with those of
Gallagher \& Hunter (1986) by sampling the B and V images through the same
apertures as their measurements. Of eight measurements four are within
 $0.02\;mag$ from their published values, three galaxies deviate by up to 
$0.2\;mag$, and for VCC1179 our value, as measured in a similar
aperture, is brighter by $0.4\;mag$
than that of Gallagher \& Hunter (1986). For this galaxy they have
 obtained a value of B--V redder
by $0.15\;mag$ than our expected result. VCC1179 is a very faint object
and it is likely that their aperture missed a significant part of the galaxy
due to a blind offset pointing. 

As for the B--V results, our expected values are redder by $0.1\;mag$ on 
average than those of Gallagher \& Hunter 1986, except for VCC1179, for
which our result is bluer, as mentioned above.
We note that the broad-band colors were obtained through comparative
photometry on CCD images, together with absolute photometry of the reference
stars and we check whether this could be due to a color term. The data of the 
reference stars are calibrated according to the standard
Johnson-Kron-Cousins bands, but the CCD photometry did not take into account
the possible difference between the CCD filters and the standard. However,
if the CCD+filters observing system has a nonzero color term, this would affect
the results only through the color {\it difference} between the measured
galaxy and the reference star on the CCD frames. In other words, if the 
galaxy is
measured to have the same CCD colors as the reference star, then its standard
colors are the same as the star's, provided the color term is
not too large. The color differences between the sample galaxies
and the stars
were mostly in the range of 0.0 -- 0.2$\;mag$ in all colors. This means that a
color term as large as 0.15 will cause an error of up to 0.03$\;mag$ in the
results, which is the typical error in our data. From a number of
observations carried out in 1993, the WiseObs 
RCA CCD was found to have color terms of up to 0.15 (in B--V, Heller 1993,
private communication). The change in 
colors due to this effect is, thus, not markedly significant in our case.
In addition, no correlation was found between the color difference of
galaxy - ref. star and the difference from published data.
Based on this consideration, we believe that our broad-band results are 
reliably representing
the observed standard color indices of the galaxies.

%
\begin{table}[htbp]
\vspace{12.9cm}
\includegraphics{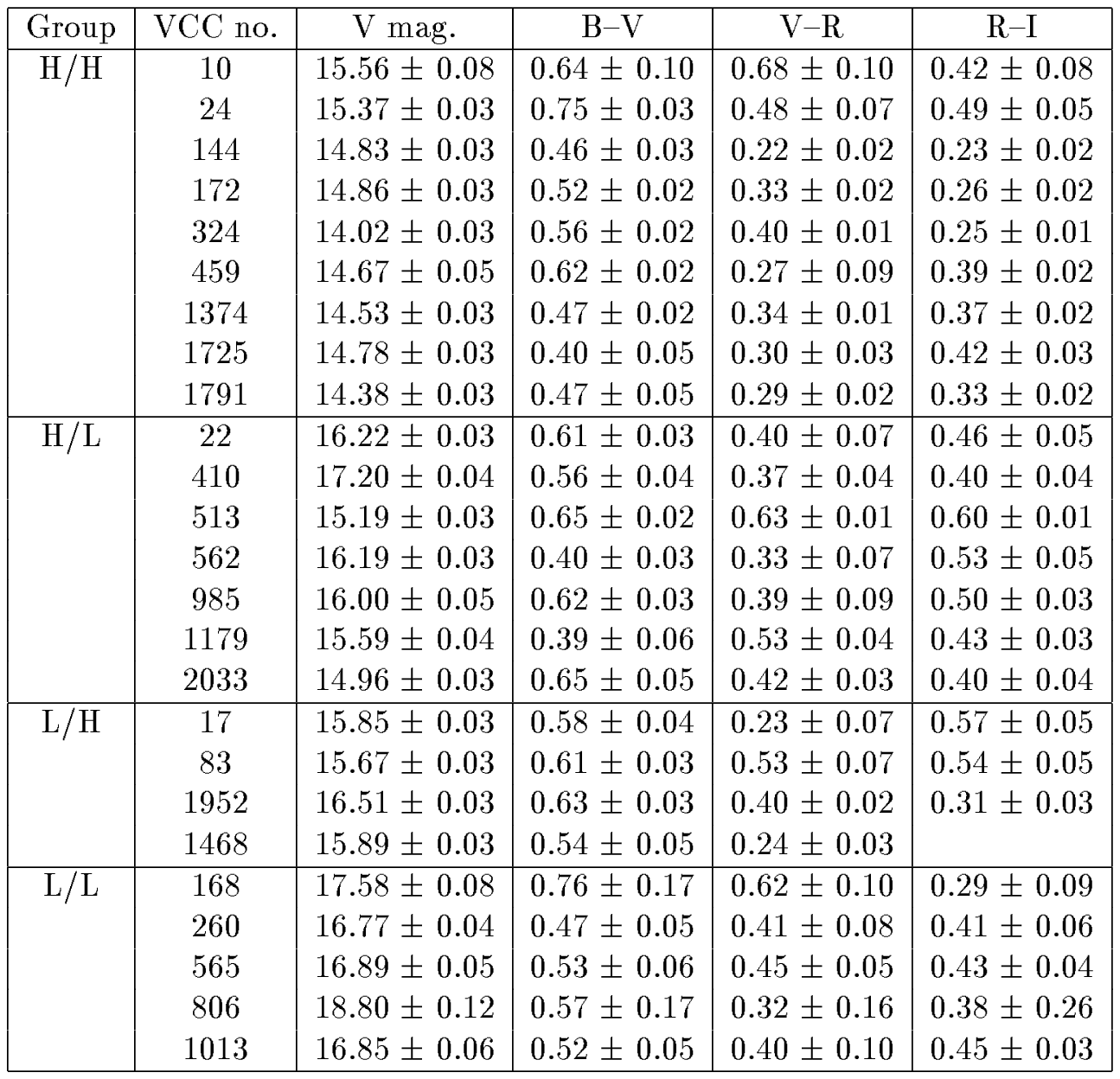}
\caption{\protect \footnotesize{Broad-band magnitudes and color indices
 of the sample.}} \label{tab_BB}

\end{table}


The broad-band data of the galaxies are given in Table~\ref{tab_BB}.
These broad-band colors are the {\it observed} values, and are not 
corrected
for internal dust extinction (the Galactic extinction is very small
in the direction of Virgo).  The issue of internal dust extinction
will be discussed in a subsequent paper.

    \subsection{H$\alpha$ fluxes} \label{sub_HaRes}

Out of 25 galaxies with broad-band data, 21 were measured in H$\alpha$, while
the remaining four were too faint to be observed through such
narrow bandpasses. 
The total H$\alpha$ + [N II] fluxes of these 21 galaxies 
are as follows: 7 objects were found to have fluxes consistent with zero
within 1.5$\sigma$, and the others range 
form $1.7\times10^{-14}$ to $7\times10^{-13} erg/cm^2/s$ with an average of
$2.3\times10^{-13} erg/cm^2/s$. The 
H$\alpha$-brightest galaxies (above $2\times10^{-13} erg/cm^2/s$) are those 
from the high surface brightness 
group. Of these, the high HI galaxies subsample seems to dominate.
However, this could be a selection effect, with the high HI galaxies
closer and larger, in general, than the low HI galaxies. These results,
together with other data, are given in Table~\ref{tab_Ha1}.
Considering the H$\alpha$ + [N II] surface brightness, rather than the total
line flux,
the two subsamples are not markedly different from each other. 
The H$\alpha$ + [N II] surface 
brightness ranges up to $2.1\times10^{-15} erg/cm^2/s/\Box"$, with
the typical value a few times smaller. Hereafter we refer to
H$\alpha$ + [N II] as 'H$\alpha$', both for convenience and, as will be
discussed below, since these objects have probably
 relatively small contributions of nitrogen lines.

As mentioned in~\ref{sub_Ha}, the filters used 
have a bandpass of $\simeq$50\AA \ and the H$\alpha$ lines deviate from
the central wavelength by at most 12\AA. 
The H$\alpha$ lines themselves have a 
typical width corresponding to a velocity dispersion of $\sim180$ km/s, 
 i.e., $\sim4$\AA. The combination of two
 profiles - the filter and the emission line makes it very difficult
to correct the measured flux. We choose, therefore, not to attempt any 
correction, knowing that the derived fluxes may be subject 
to a reduction of typically 5 percent due to this effect.

Another correction that cannot be accurately derived without detailed 
spectroscopic information is the contribution of the  
nitrogen lines to the measured H$\alpha$ line. This is mainly
because galaxy lines are shifted differently relative
to the peak transmission wavelength of the filters and the [N II] lines are
sampled through different parts of the filters' response curve for every 
galaxy. As a consequence, the [N II] lines are weighted 
differently relative to the H$\alpha$ line for each galaxy. However,
neglecting the [N II] 
contribution will not affect our results significantly.
 The typical [N II]/H$\alpha$ ratio is less than 
0.10 for dwarf irregulars (e.g., Kennicutt 1983, Gallagher \& Hunter 1989), 
as in general
they have lower metallicities than spirals and lenticulars. This effect
is in the opposite direction to the previous one, namely, it increases the
flux we assign to the H$\alpha$ line, while the effect of missalignment of 
the line with the filter decreases the measured flux.
To our estimate, both effects are within the errors of our 
measurements.

A greater effect on the H$\alpha$ results is the internal dust extinction.
Again, this is different from one galaxy to another, and is more difficult
to account for. 
As in the broad-band section, we stick to the raw data 
as far as presentation of the results is concerned and consider this 
effect in
a subsequent paper. This applies for both the H$\alpha$ fluxes and 
 the equivalent widths of the lines. The H$\alpha$ equivalent widths are 
also shown in Table~\ref{tab_Ha1}, together with the corresponding fluxes.

\begin{table}[htbp]
\vspace{11.2cm}
\includegraphics{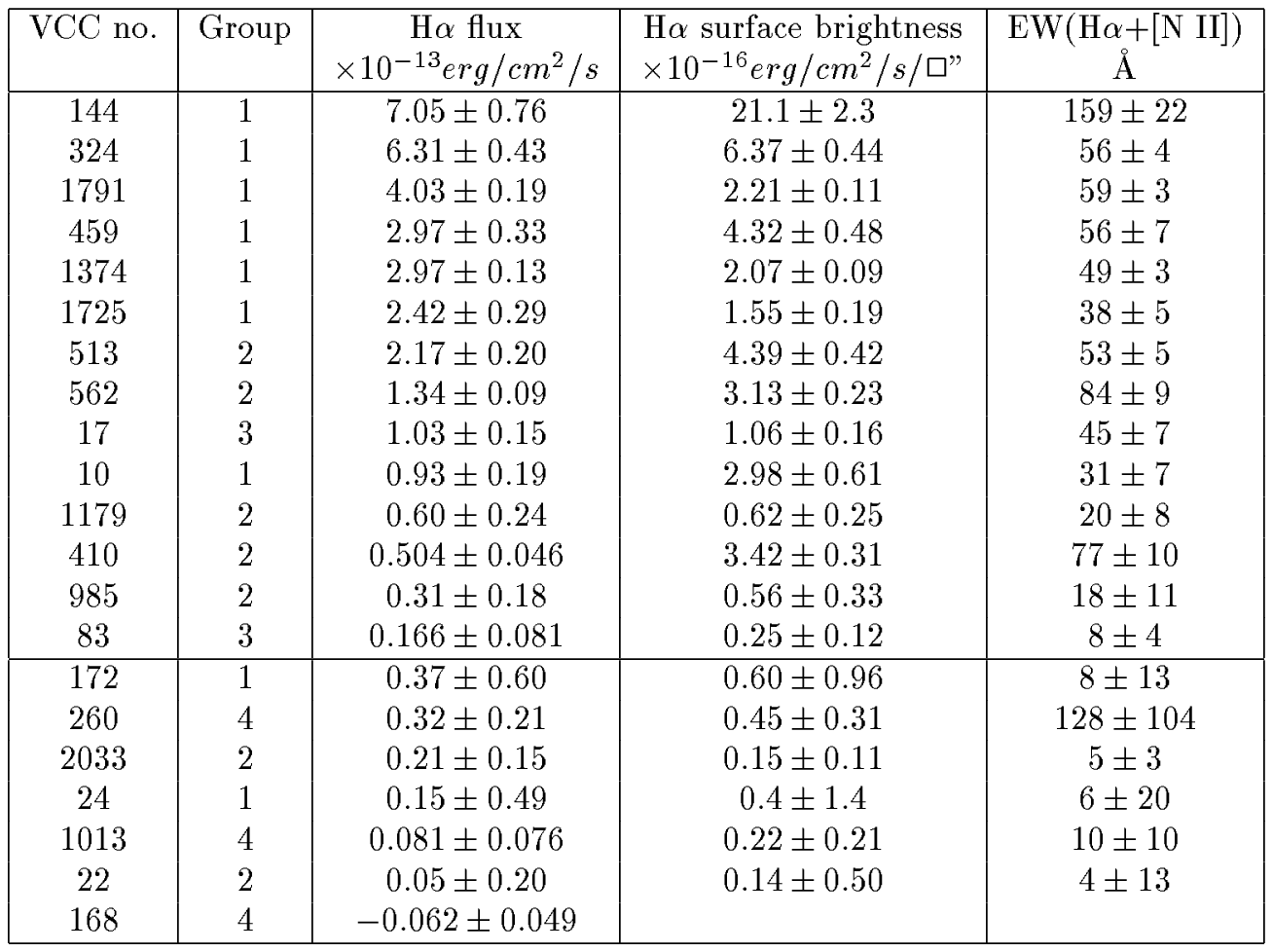}
\caption{\protect \footnotesize{H$\alpha$ data in 
decreasing order of H$\alpha$ line flux. The galaxies in the
lower panel have values consistent with zero within 1.5$\sigma$.}} 
\label{tab_Ha1}

\end{table}

    \subsection{Combined results} \label{sub_com}

\begin{figure}[htbp]
\vspace{7.5cm}
\includegraphics{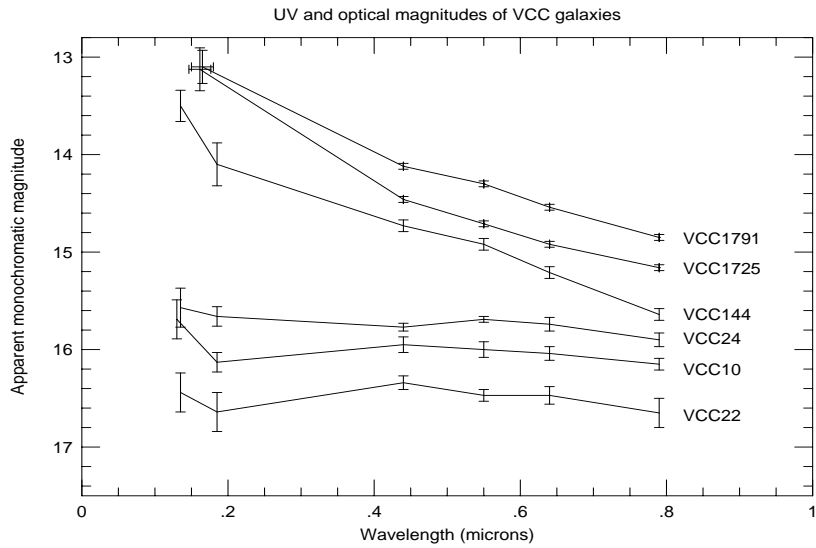}
\caption{\protect \footnotesize{Spectral energy distribution of sample galaxies 
with UV data. The UV points of VCC10 and VCC1725 are slightly shifted
for clarity.}}
\label{fig_sed1}
\end{figure}

\begin{figure}[htbp]
\vspace{7.5cm}
\includegraphics{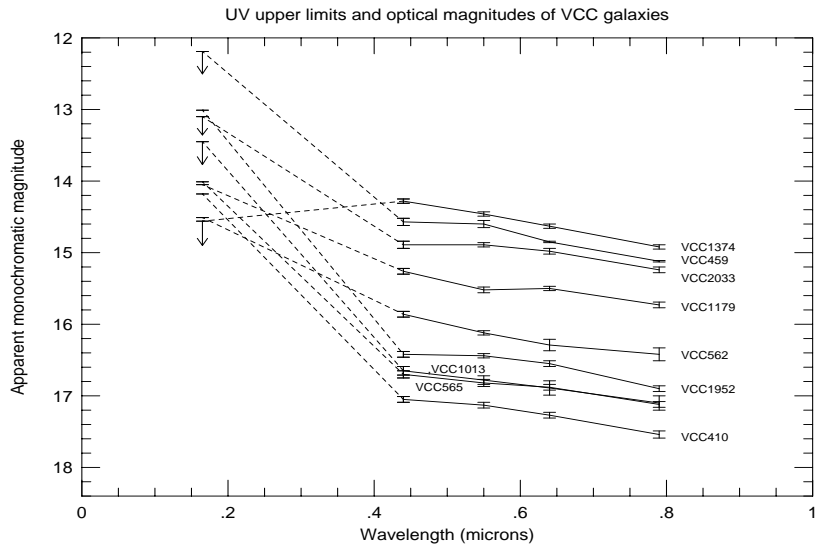}
\caption{\protect \footnotesize{Spectral energy distribution of sample galaxies 
which have
upper limits for UV radiation. For clarity, not all upper limit arrows
are shown.}}
\label{fig_sed2}
\end{figure}

The UV monochromatic magnitudes obtained from the IUE observations
or from FAUST were combined with the optical monochromatic magnitudes
from WiseObs to yield a rough spectral energy distribution
of the sample galaxies. No correction for internal extinction was made.
Unfortunately, only 6 galaxies were detected in the UV. Nine other
galaxies
have UV upper limits from FAUST images and nine others have no UV 
data at all. The galaxies with UV magnitudes are plotted in 
Fig.~\ref{fig_sed1} and
 the galaxies with upper limits in Fig.~\ref{fig_sed2}.

As expected from the broad-band data, the SEDs of the galaxies display a 
variety of
behaviors. In Fig.~\ref{fig_sed1}, three of the galaxies have steep
SEDs (two were detected by FAUST and VCC144 was chosen for observation by
IUE due to its strong H$\alpha$ emission), and three have flat SEDs.
It should be noted that for galaxies with IUE data only the SED of
the inner part of the galaxy is plotted.

     \section{Discussion} \label{sec_dis}

\begin{figure}[htbp]
\vspace{7.5cm}
\includegraphics{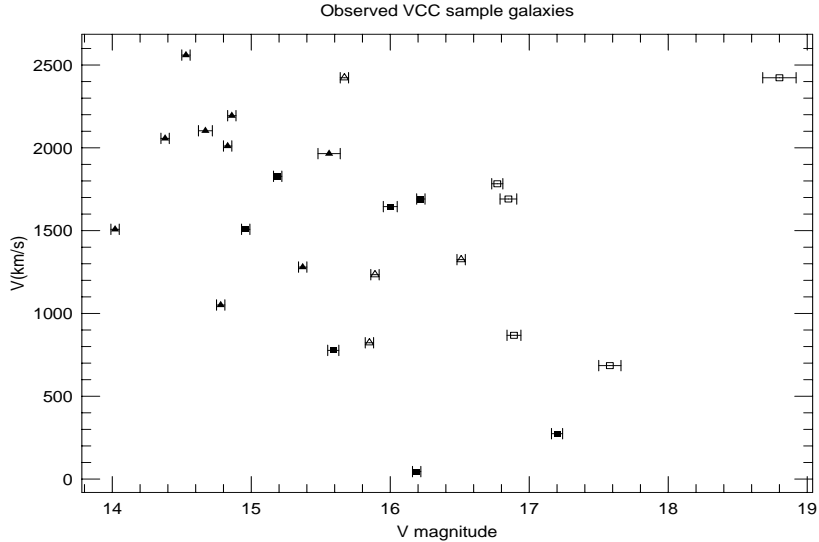}
\caption{\protect \footnotesize{Recession velocities of the sample galaxies 
versus the observed
magnitude. The negative slope may indicate infall of the galaxies into
the cluster core (see text).}}
\label{fig_Vv}
\end{figure}

   An interesting relation is apparent between the V magnitude and the
   recession velocity of the galaxies, as derived from the HI line
   (Fig.~\ref{fig_Vv}). There is a general negative correlation between
the two parameters. The point in the upper right corner of the plot is
VCC806, a very faint object barely detected in the radio and optical, which
may be a background galaxy. If we exclude this point,
the correlation coefficient is --0.55. The correlation coefficient between
the recession velocity and the V-band fluxes of the galaxies corresponding to 
their magnitudes is 0.46.

This correlation may lead to 
the following empirical conclusion: if faint galaxies in Virgo are 
characterized by lower recession velocities,
a Malmquist bias will cause the depletion of slowly receding galaxies 
in the sample. This may explain the observed skewness of the velocity 
distribution of faint dwarfs in Virgo (Binggeli {\it et al.} 1993).
 These velocities are shifted, in general, towards higher values, 
compared with the velocity distribution of
other galactic types. We may expect, therefore, that
many faint dwarf galaxies which do not yet have velocity data are mostly 
 slowly receding, with velocities in the range of 0--1500 km/s. This may
balance the skewness of the velocity distribution of these galaxies.

Clearly, checking this relation with larger samples in Virgo,
not necessarily of faint blue galaxies is necessary, but such a survey is 
outside the scope of this paper.

As for the origin of this correlation, it may arise because of
 infall of galaxies towards the Virgo cluster core (e.g., Tully \& 
Shaya 1984, Binggeli {\it et al.} 1993). Galaxies which are between 
us and the core appear brighter and are falling
away from the observer to the cluster core, while galaxies behind 
the core are falling
`toward' us, having a recession velocity lower than that
of the core itself (around 1200 km/s). This would produce the effect of
negative correlation seen here and, according to this interpretation, the 
faint Virgo
dwarfs lacking velocity data would be those behind the cluster core.
 It should be noted, however, that this effect, in its simple interpretation,
 does not match the data perfectly. This is because the range of
magnitudes of the sample galaxies, which is correlated to the recession
velocity, is roughly 1.5 -- 3 $mag$, and depends on the intrinsic scatter
of absolute magnitudes assumed for the sample.
This range corresponds to a difference in distance of a factor of 2 -- 4 
between
the near and far galaxies of the sample, which is not likely to be the case in
Virgo. 

\begin{figure}[htbp]
\vspace{14.8cm}
 \includegraphics{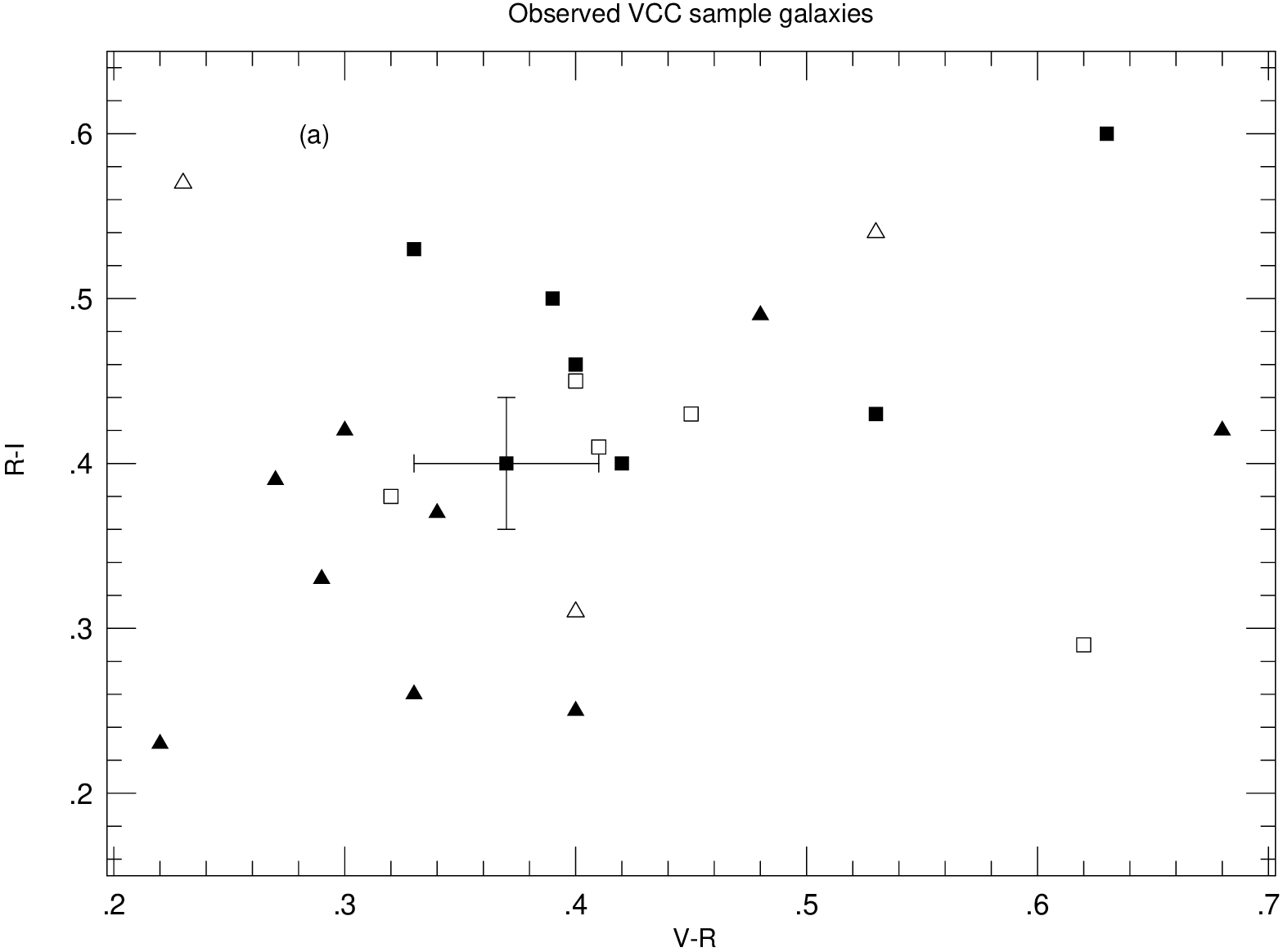}
\includegraphics{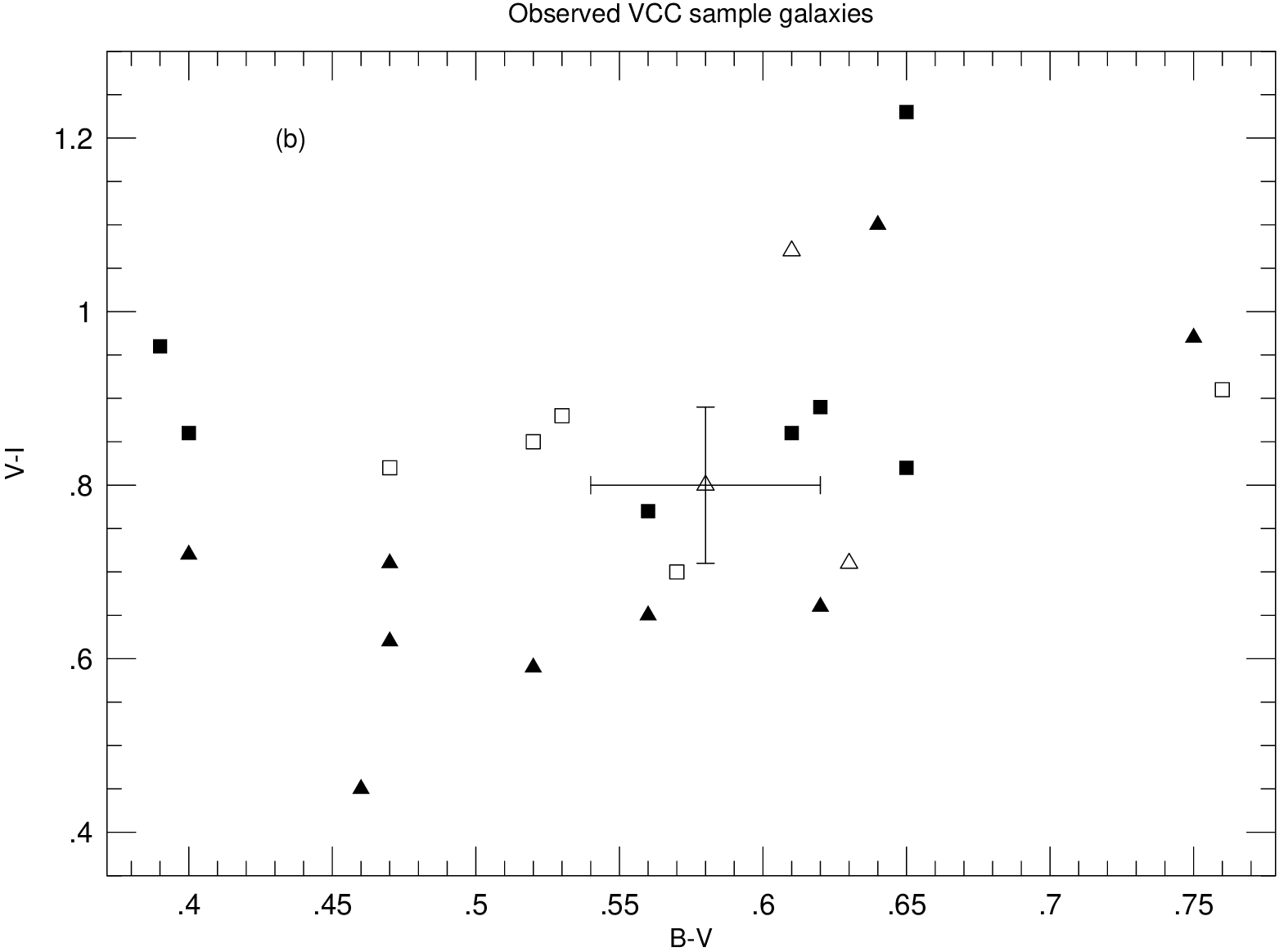}
\caption{\protect \footnotesize{Color-color diagrams of the 
sample galaxies. In (a) the R band 
increases the dispersion of the objects. In (b) this band is not included
and the dispersion is smaller. The galaxies with high HI 
content are marked
by triangles and those with low HI content by squares. 
High surface brightness galaxies are marked by filled symbols and the
low surface brightness ones by empty symbols. A typical error bar is displayed 
with one of the points in each diagram.}}
\label{fig_BRI1}
\end{figure}

    The broad-band B--V colors of the sample galaxies are relatively blue,
 ranging from $\sim\!0.42$ to $\sim\!0.72$.
 Unlike our previous assumption presented
in section~\ref{sec_SAM},
no correlation between B--V and the optical surface brightness
 was found. The scatter in the B--V results indicates that
the galaxies do not share the same properties. This is better demonstrated
in the color-color diagram of V--R vs. R--I [Fig.~\ref{fig_BRI1}a]. The 
error bar shown there is a typical one, but the points that deviate significantly
from the general locus (in the lower right and upper left of the figure)
have error bars up to 2.5 times larger that this.

In this
 diagram, the points are further dispersed due to the variation of
H$\alpha$ equivalent widths of the galaxies. The H$\alpha$ line 
is included in the broad R band and may change the total R flux
of strong-emission-line galaxies.
 For this reason, the scatter is smaller in the B--V vs. V--I diagram,
which is shown in Fig.~\ref{fig_BRI1}b. However, note that in the galaxy
with the most intense H$\alpha$ (VCC144), the contribution of the H$\alpha$
line to the broad-band R is less than 15\%. As in Fig.~\ref{fig_BRI1}a, the
errors of the deviating points here are also larger than the typical ones.
Another reason for the scatter
may be different amounts of internal extinction in each galaxy, 
which may arbitrarily
shift the galaxy on the diagram, albeit in the general direction of the
color-color dependence.

In Figs.~\ref{fig_BRI1}a and~\ref{fig_BRI1}b the galaxies with high HI 
content are marked
by triangles and those with low HI content by squares. 
High surface brightness galaxies are marked by filled symbols and
low surface brightness ones by empty symbols. 
There are clearly not enough low
surface brightness objects for meaningful statistical conclusions, but
the high surface brightness group shows a difference between the high HI
content (filled triangles) and the low HI content (filled squares). They
share the same range of B--V color index but the lower HI group seems to
have redder V--R color. The sample should be enlarged before 
clearer conclusions can be drawn about this effect.

The measured H$\alpha$ equivalent widths of the galaxies
 are relatively high, typically 
$\sim$50\AA,
and up to 159\AA. Again, the high surface brightness sub-samples have higher
 EW[H$\alpha$] than others. This implies that a large fraction 
of the surface of the galaxies is covered by
one or more HII regions. Some BCD galaxies are also known as extragalactic HII
regions or HII galaxies. This name is very suitable for some 
of the sample galaxies. 
In the case of VCC144, where the equivalent width
is 159\AA, the entire galaxy is one giant HII region (Brosch {\it et al.} 1997b). 

The high H$\alpha$ flux and EW[H$\alpha$], together with the spatial
appearance of the sample galaxies in H$\alpha$, namely the fraction
of galactic surface covered by HII regions, emphasize
 the high SFR of some of the galaxies and the starburst activity that takes 
place in them. 
On the other hand, other galaxies
in the sample show little or no H$\alpha$ emission, which indicates
their low current star formation activity. There appears to be no correlation
between the H$\alpha$ flux and the HI 21cm line flux.

\section{Conclusion} \label{sec_conc}  

We presented observations related to star formation properties of
a sample of late-type dwarf galaxies. 
The intention in focusing on dwarf galaxies 
was to exclude some of the star formation inducing mechanisms,
 assumed to account for
star formation in large galaxies. In addition, we concentrate on Virgo
cluster members, in order to test the effects of the cluster 
environment on the star formation properties of the galaxies.

 A data base consisting of a number of broad-band colors and
H$\alpha$ line observations is important for determining
the ongoing star formation process, as well as the star formation history of
the sample. In a subsequent paper we will show
 that some of the galaxies show signs of a strong burst
of star formation, while others completely lack signs of recent
 star formation activity.

The observational data are affected
primarily by internal dust extinction in the galaxies, which
 complicates the interpretation of the data. This effect
 will be discussed in detail in the next paper.
 
Another interesting finding, concerning the dwarf galaxies in Virgo, 
is their velocity 
field. Our results suggest an infall of the galaxies towards the cluster 
core, which may explain partly the skewed
recession velocity distribution of the dwarf galaxies, in terms of a
Malmquist bias.

\section*{Acknowledgments}

Observations at the Wise Observatory are partly supported by a Center of 
Excellence Grant from the Israel Academy of Sciences.
UV studies at the Wise Observatory are supported by special grants from
the Ministry of Science and Arts, through the Israel Space Agency, to
develop TAUVEX, a UV space imaging experiment, and by the Austrian Friends
of Tel Aviv University. 

EA was supported partly by a grant from "The Fund for the
Encouragement of Research" Histadrut- The General federation of Labour in
Israel.
NB acknowledges the hospitality of Prab Gondhalekar
and of the IRAS Postmission Analysis Group at RAL, as well as IRAS Faint
Source catalog searches by Rob Assendorp.

We thank Stuart Bowyer and Tim Sasseen from the Space Sciences Laboratory, 
Berkeley, University of California,
 for kindly providing the FAUST images of the Virgo cluster.

\section*{References}
\begin{description}

\item{Almoznino. E., Loinger, F. \& Brosch, N. 1993, Mon. Not. R. astr. Soc. \mbox{{\bf 265}, 641.}}

\item{Binggeli, B., Sandage, A. \& Tammann, G.A. 1985, Astron.J. {\bf 90}, 
1681 (BST).}
 
\item{Binggeli, B., Popescu, C.C. \& Tammann, G.A. 1993, Astron. Astrophys. Suppl. 
 {\bf 98}, 275.}
 
\item{Bowyer, S., Sasseen, T.P., Wu, X. \& Lampton, M. 1995, 
Astrophys. J. Suppl. {\bf 96}, 461.}

\item{Brosch, N., Almoznino, E., Leibowitz, E.M., Netzer, H.,
 Sasseen, T.P., Bowyer, S., Lampton, M. \& Wu, X. 1995, Astrophys. J. {\bf 450}, 137.}

\item{Brosch, N., Formiggini, L., Almoznino, E., Sasseen, T.P., Lampton, M. 
 \& Bowyer, S. 1997a, Astrophys. J. Suppl. {\bf 111}, 143.}

\item{Brosch, N., Almoznino, E. \& Hoffman, G.L. 1997b, Astron. Astrophys. {\it in press}.}


\item{de Vaucouleurs, G., de Vaucouleurs, A., Corwin, H.G. Buta, R.J.,
Paturel, G. \& Fouqu\'{e}, P. 1991, {\it The Third Reference Catalog of
Bright Galaxies}, (Springer, New-York).}
 
\item{Drinkwater, M. \& Hardy, E. 1991, Astron.J. {\bf 101}, 94.} 

\item{Fouqu\'{e}, P., Bottinelli, L., Gouguenheim, L. \& Paturel, G. 1990, 
Astrophys. J. \mbox{{\bf 349}, 1.}}

\item{Gallagher, J.S., Hunter, D.A. \& Tutukov, A.V. 1984, Astrophys. J. {\bf 284}, 544 (GHT).}
 
\item{Gallagher, J.S. \& Hunter, D.A. 1986, Astron.J. {\bf 92}, 557.}
 
\item{Gallagher, J.S. \& Hunter, D.A. 1989, Astron.J. {\bf 98}, 806 (GH89).}

\item{Gallego J., Zamorano J., Rego M., Alonso O. \& Vitores A.G. 1996, Astron. 
 Astrophys. Suppl. Ser. {\bf 120}, 323.}

\item{Hoffman, G.L., Helou, G., Salpeter, E.E., Glosson, J. \& Sandage, A.
 1987,  Astrophys. J. Suppl. {\bf 63}, 247.}

\item {Hoffman, G.L., Williams, H.L., Salpeter, E.E., Sandage, A. \&
Binggeli, B. 1989, Astrophys. J. Suppl. {\bf 71}, 701.}

\item{Hunter, D.A. \& Gallagher, J.S. 1985, Astron. Astrophys. Suppl. 
{\bf 58}, 533.}

\item{Kennicutt, R.C. 1983, Astrophys. J. {\bf 272}, 54 (K83).}
 
\item{Kennicutt, R.C. \& Kent, S.M. 1983, Astrophys. J. {\bf 88}, 1094.}
 
\item{Kennicutt, R.C. 1989, {\it Large Scale Star Formation \& the Interstellar
 Medium}, in "The Interstellar Medium in External Galaxies". ed.
 H.A.Thronson \& J.M.Shull.}
  
\item{Kennicutt, R.C., Tamblyn, P. \& Congdon, C.W. 1994,  Astrophys. J. 
{\bf 435}, 22.}
 
\item{Kunth, D. \& Sargent, W.L.W. 1986, Astrophys. J. {\bf 300}, 496.}

\item{Landolt, A.U. 1973, Astron.J. {\bf 78}, 958.}

\item{Landolt, A.U. 1992, Astron.J. {\bf 104}, 340.}

\item{Massey, P., Strobel, K., Barnes, J.V. \& Anderson, E. 1988,  Astrophys. J. 
\mbox{{\bf 328}, 315.}}

\item{Pogge, R.W. \& Eskridge, P.B. 1987, Astrophys. J. {\bf 93}, 291 (PE87).}

\item{Pogge, R.W., Goodrich, R.W. \& Veilleux, S. 1988, {\it The VISTA Cookbook},
Lick Observatory Technical Report No. 50, University of California, CA.}

\item{Shemi, A., Mersov, G., Brosch, N. \& Almoznino, E. 1993, {\it The 
Prediction of Stellar Ultraviolet Colours}, in "Astronomical Data
Analysis Software and Systems III", ed. Crabtree, D.R., Hamisch, R.J.
and Barnes, J. (A.S.P. conference series, Volume 61).}

\item{Tully, R.B. \& Shaya, E.J. 1984, Astrophys. J. {\bf 281}, 31.}
 
\item{van der Hulst, J.M., Kennicutt, R.C., Crane, P.C. \& Rots, A.H. 1988, 
Astron. Astrophys. {\bf 195}, 38.}

\end{description}


\end{document}